\documentclass[amsmath,amssymb,pra,superscriptaddress,twocolumn]{revtex4-2}
\usepackage{graphicx,url,siunitx}
\usepackage[colorlinks=true,breaklinks=true]{hyperref}
\usepackage{color,physics,cleveref,soul}
\usepackage[english]{babel}

\DeclareSIUnit\gauss{\text{G}}

\graphicspath{{figures/}}

\begin{document}
\title{Loading atoms from a large magnetic trap to a small intra-cavity dipole trap}
\author{D. Varga}
\affiliation{HUN-REN Wigner RCP, H-1525 Budapest P.O. Box 49, Hungary}
\affiliation{Department of Physics of Complex Systems, ELTE Eötvös Loránd University, Pázmány Péter sétány 1/A, H-1117 Budapest, Hungary}
\author{B. Gábor}
\affiliation{HUN-REN Wigner RCP, H-1525 Budapest P.O. Box 49, Hungary}
\affiliation{Department of Theoretical Physics, University of Szeged, Tisza Lajos k\"{o}r\'{u}t 84, H-6720 Szeged, Hungary}
\author{B. Sárközi}
\affiliation{HUN-REN Wigner RCP, H-1525 Budapest P.O. Box 49, Hungary}
\author{K. V. Adwaith}
\affiliation{HUN-REN Wigner RCP, H-1525 Budapest P.O. Box 49, Hungary}
\author{D. Nagy}
\affiliation{HUN-REN Wigner RCP, H-1525 Budapest P.O. Box 49, Hungary}
\author{A. Dombi}
\affiliation{HUN-REN Wigner RCP, H-1525 Budapest P.O. Box 49, Hungary}
\author{T. W. Clark}
\affiliation{HUN-REN Wigner RCP, H-1525 Budapest P.O. Box 49, Hungary}
\author{F. I. B. Williams}
\affiliation{HUN-REN Wigner RCP, H-1525 Budapest P.O. Box 49, Hungary}
\author{P. Domokos}
\affiliation{HUN-REN Wigner RCP, H-1525 Budapest P.O. Box 49, Hungary}
\author{A. Vukics}
\affiliation{HUN-REN Wigner RCP, H-1525 Budapest P.O. Box 49, Hungary}

\begin{abstract}
We show that an optimized loading of a cold ensemble of rubidium-87 atoms from a magnetic trap into an optical dipole trap sustained by a single, far-red-detuned mode of a high-Q optical cavity can be efficient despite the large volume mismatch of the traps. The magnetically trapped atoms are magnetically transported to the vicinity of the cavity mode and released from the magnetic trap in a controlled way meanwhile undergoing an evaporation period. Large number of atoms get trapped in the dipole potential for several hundreds of milliseconds. We monitor the number of atoms in the mode volume by a second tone of the cavity close to the atomic resonance. While this probe tone can pump atoms to another ground state uncoupled to the probe, we demonstrate state-independent trapping by applying a repumper laser.
\end{abstract}

\maketitle

\clearpage

\section{Introduction}
The manipulation of neutral atoms by light fields \cite{chu1998nobel,cohen1998nobel,phillips1998nobel} has become a major domain within experimental quantum optics research during the past half a century. Cooling and trapping single neutral atoms and molecules or (ultra)cold ensembles thereof, and guiding beams composed of such particles \cite{cornell2002nobel,ketterle2002nobel} have led to an immense variety of experiments. Today, such systems provide one of the major pillars of quantum science with applications both in quantum information \cite{porto2003quantum,schrader2004neutral} and quantum metrology \cite{pezze2018quantum}, and spin-offs in quantum technology.

Cavity quantum electrodynamics has emerged as a platform for the study of light-matter interaction at its purest, since individual electromagnetic modes can be made interact with carefully selected atomic transitions. Whereas in a strong-coupling regime, the cavity modes can be present as dynamical elements \cite{ritsch2013cold,dombi2021collective,clark2022time,gabor2023ground,suarez_collective_2023,rivero_quantum_2023}, cavities are routinely used merely for enhanced dipole potentials \cite{edmunds_deep_2013,bothwell2022resolving}, or increased sensitivity of atom detection \cite{greve2022entanglement}. In a cavity QED experimental system, besides controlling the atom-photon interaction, the most advanced trapping and manipulation techniques are needed to keep the atoms for long-enough time within the tiny interaction volume of an optical resonator.

Purely magnetostatic trapping of neutral laser-cooled ensembles \cite{migdall1985first,fortagh2007magnetic} is a convenient way to manipulate the spatial characteristics of the cloud, as such traps can be easily shaped and displaced \emph{in situ}. The prototype is the magnetic quadrupole trap \cite{bergeman1987magnetostatic}, where the relative currents in a pair of coils determines the trap size and position. However, such traps rely on an adiabatic potential, that involves careful state preparation within the hyperfine structure by optical pumping. Very cold clouds can violate the adiabaticity condition, e.g. a Bose-Einstein condensate cannot be stored in a magnetic quadrupole trap due to the loss of atoms in the zero-field center. Furthermore, the perturbations of the atomic levels by the trapping field can be prohibitive in many situations.

Optical dipole traps \cite{grimm2000optical} have formed another route towards spatial manipulation of cold and ultracold atomic clouds. Composed of very far detuned optical fields, they can be considered largely state-independent conservative traps for many applications. However, since the trap depth depends on the intensity profile, the spatial manipulation of such traps is more cumbersome. A spectacular example is the optical conveyor belt \cite{schrader2001optical}.


In this work, we connect the above tendencies by demonstrating the loading of a cavity-sustained one-dimensional optical dipole trap lattice \cite{kruse_cold_2003,mckeever_state-insensitive_2003,maunz_normal-mode_2005,puppe2007trapping,baumann_dicke_2010,roux_cavity-assisted_2021} from a large  ensemble of cold rubidium-87 atoms transported in a quadrupole magnetic trap. The loading of atoms directly from a magnetic into the micron-sized cavity dipole trap is a highly non-trivial task because of the large size mismatch. So far, either a free-space optical lattice \cite{sauer_cavity_2004,nussmann_submicron_2005,brennecke_cavity_2007} or tweezer \cite{deist_mid-circuit_2022} was used to load from a large magnetic trap, or an atom chip with a small magnetic trapping volume was utilized to guide atoms into interaction with a cavity mode \cite{teper_resonator-aided_2006,trupke_atom_2007,murch_observation_2008}.
In the present work, we show that the routinely used bulky free-space magnetic traps can also achieve this task, provided that the process is carefully optimized, when the goal is to have many atoms in a relatively large, centimeter long open resonator.

The overall experimental cycle is the following. A cold $^{87}$Rb atomic sample is prepared in a magneto-optical trap above the cavity axis by about $\SI{1.1}{\cm}$. Upon polarization-gradient cooling we apply optical pumping to the $F=2,\,m_F=2$ ground state, in order to load the atoms into a quadrupole magnetic trap. By varying the currents in the coil pair, we displace the center of the trap vertically to the vicinity of the cavity axis. The atomic cloud follows adiabatically, and gets released from the magnetic trap in a controlled ramp down of the currents. A fraction of the atoms is captured by the intra-cavity dipole trap created in one of the far-red-detuned cavity modes. We optimize several parameters of this process in order to maximize the captured atom number.

The intra-cavity atom number is determined by measuring the transmission of another laser-driven mode of the cavity close to the atomic resonance. This mode undergoes significant dispersive shift that depends on the number of atoms interacting with the mode. This measurement has a back-action on the trapping: besides heating the atoms, the probe laser optically pumps the atoms into the other hyperfine ground state, where their dispersive effect on the cavity is negligible (depumping). Hence, the atoms are “tucked away” to the $F=1$ ground state, where they sit in the dark, feeling only the dipole trap potential. However, they can be pulled back to $F=2$ by the application of a third laser, the repumper, so that their effect on the cavity reappears. Via this procedure, we observe trapping times up to several hundreds of milliseconds in the cavity mode.

\section{System and experimental protocol}

\begin{figure*}
 \includegraphics[width=.9\linewidth]{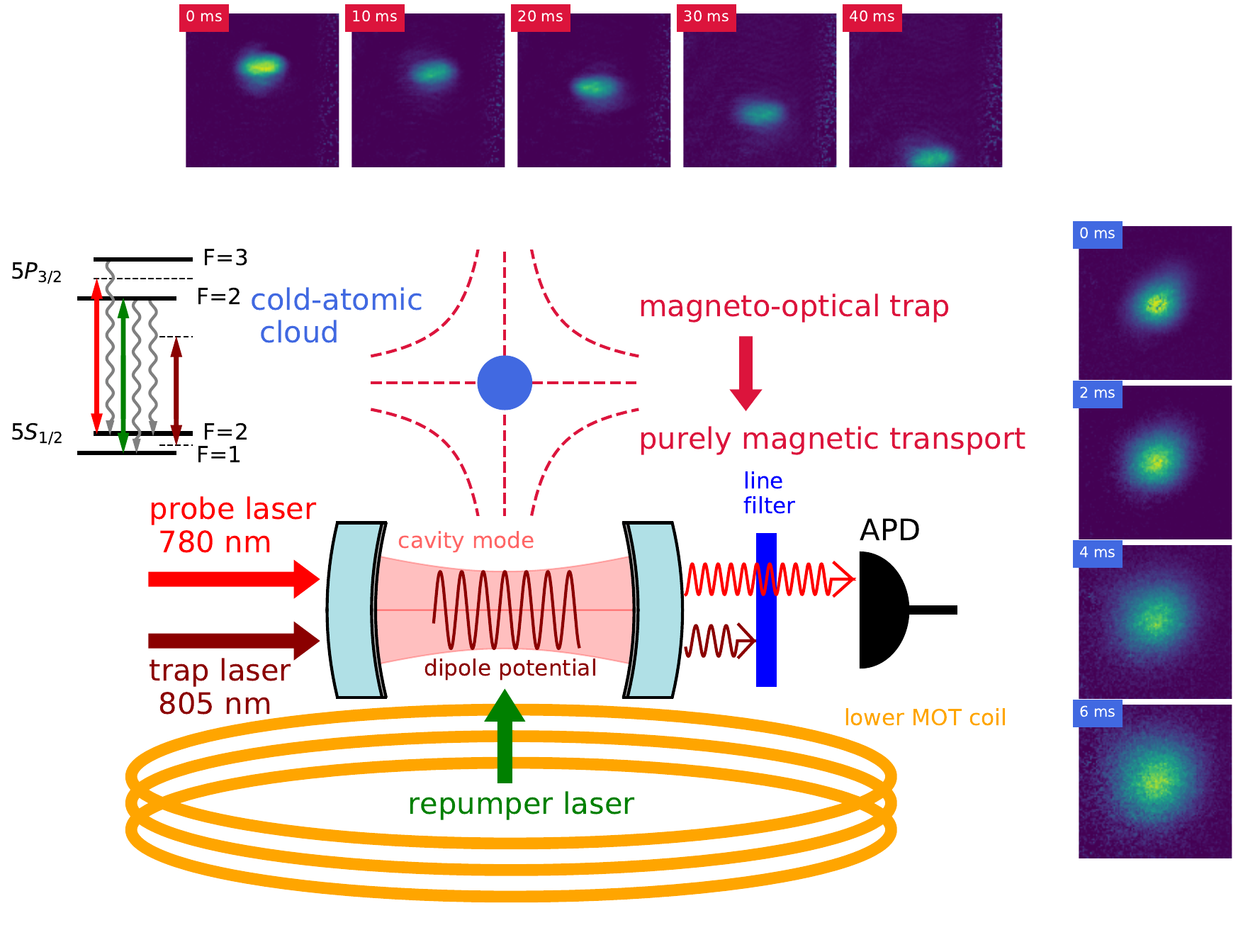}
 \caption{Schematic of the experimental apparatus. A cold ensemble of $^{87}$Rb is created by magneto-optical trapping and polarization-gradient cooling. After optical pumping into trapped states in a magnetic quadrupole configuration, the cloud is magnetically transported downwards into the optical cavity (illustrated by the horizontal sequence of absorption images). The cavity has two modes relevant to the experiment: (1) the dipole-trapping mode far-red-detuned ($\approx\SI{805}{\nano\meter}$, dark red arrow) from the atomic resonance (2) a mode close to the atomic resonance (red arrow) used for monitoring the intra-cavity atom number via the dispersive effect of the atoms. On the cavity output, the trap light is filtered out, while the intensity of the probe light is monitored by an avalanche photodetector. To recover the atoms depumped into the hyperfine ground state uncoupled to the probe, we can optionally apply a repumper tone (green arrow). The relevant atomic levels and the laser tones are depicted in the top left corner of the figure, using the same color code for the latter as mentioned above. The vertical sequence of absorption images shows the expansion of the atomic cloud upon switching off the magnetic trap.}
 \label{fig:scheme}
\end{figure*}

The experimental scenario is depicted in \cref{fig:scheme}. An ensemble of $^{87}$Rb atoms are collected and cooled in a magneto-optical trap (MOT) using the D$_2$ line. The base pressure of the ultra-high-vacuum chamber is $\sim10^{-11}\,\text{mbar}$. Our system of electromagnets is a derivative of the one presented in \cite{cano2011experimental}, with intra-vacuo coils allowing for small drive currents $\lesssim\SI{5}{\ampere}$ that we produce by HighFinesse BCS devices.
We use an AS-Rb-0100-3V AlfaSource Rubidium dispenser with its opening oriented towards the center of the MOT. We use 6 independently adjustable MOT beams derived from a Toptica TA Pro laser. The repumper is a DFB laser built on campus, whose light is injected from top and bottom using the same vertical beam paths. We collect atoms in the MOT for $10–\SI{30}{\s}$, reaching an atom number up to $10^6$.

Following the MOT phase, we quickly ramp down the magnetic quadrupole in order to apply polarization-gradient cooling. Over \SI{5}{\ms} we reach a temperature of 20-\SI{50}{\micro\K}. The atom number and the temperature are measured by absorption imaging utilizing an IDS UI-3240ML-NIR-GL camera, cf. \cref{fig:scheme}. We note that magnetic fields are detrimental to sub-Doppler cooling by polarization gradients \cite{walhout1992sigma}, so for the efficiency of this stage, we have to carefully compensate the magnetic field of the Earth and any further stray fields surrounding the chamber.

Subsequently, over an interval of $\SI{40}{\us}$, the atoms are optically pumped into the $F=2,\,m_F=2$ ground state so that their magnetic dipole moment aligns with an applied small homogeneous magnetic field which defines the quantization axis. Both the optical pumping beam (together with the absorption imaging beam) is derived by acousto-optical modulation from a Toptica DL Pro laser used as a reference that is directly locked to rubidium reference by frequency-modulated spectroscopy.

The small homogeneous field is then adiabatically ramped into a quadrupole over a period of $\SI{3}{\ms}$ to strength $\SI{66}{\gauss\per\cm}$. The atomic dipoles adiabatically follow the direction of the local magnetic field over this process. For the subsequent magnetic transport, we ramp the currents in the coils asymmetrically to lower the zero-field center of the trap towards the cavity axis. We ramp the coil currents according to a smooth function (tangent hyperbolic), in order to avoid sudden jerks (see the nonlinear displacement of the cloud in time in \cref{fig:scheme}, horizontal absorption-image sequence). Since the vertical distance is $\sim\SI{11}\mm$ and the separation of the MOT coils is $\SI{34}\mm$, the transport brings the cloud close to the lower MOT coil. The shape of the magnetic quadrupole would therefore be strongly distorted if we used only a single coil pair. To compensate, we have built in a second coil pair of identical shape, coaxial with and situated just outside of the MOT coils, which we drive appropriately to minimize the distortion of the trap during the transport.

\begin{table}
\begin{tabular}{l|l|l}
 linewidth (HWHM) & $\kappa/2\pi$ & \SI3{\mega\hertz}\\
 finesse & $\mathcal{F}/\pi$ & 1060 \\
 length & $d$ & \SI{1.5}{\cm}\\
 free spectral range & $\nu_{\mathrm{FSR}}$ & \SI{10}{\giga\hertz}\\
 mode waist & $w_0$ & \SI{127}{\um}\\
 atom-mode coupling & $g/2\pi$ & \SI{0.33}{\mega\hertz}
\end{tabular}
 \caption{Cavity parameters. The arrangement of the mirrors is plano-spherical, so that the mode waist is situated on the planar mirror that we use for incoupling. We use LaserOptik HR mirrors of $\SI5\mm$ diameter. The reflectivity of both mirrors is $R=0.9993$, and is closely constant over a range of $\sim\SI{50}{\nm}$ around $\SI{790}{\nm}$, so the cavity parameters are the same for the two modes at $\SI{780}{\nm}$ and $\SI{805}{\nm}$ that we use in this work.}
 \label{tab:cavity}
\end{table}

\begin{figure*}
 \includegraphics[width=0.8\linewidth]{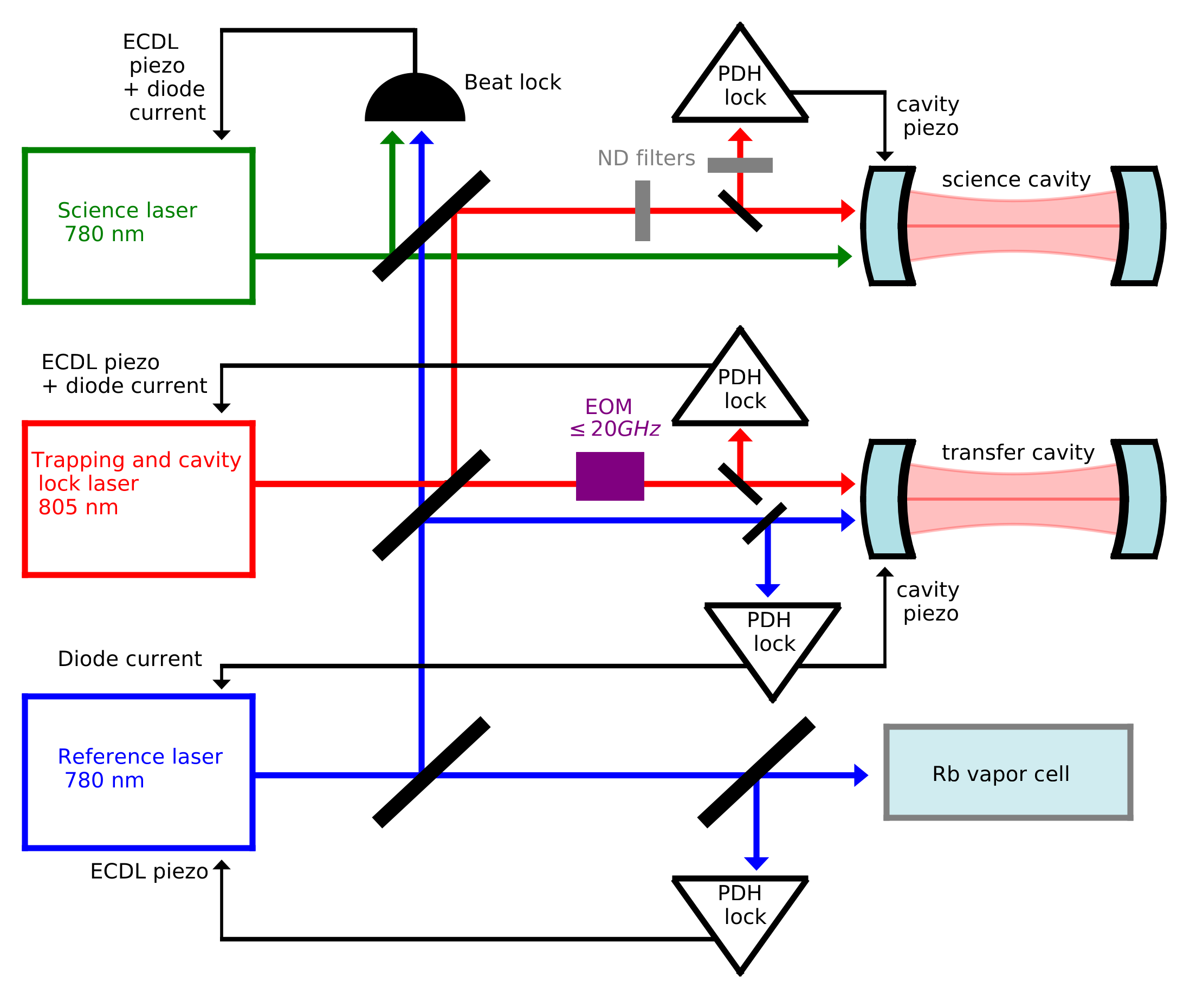}
 \caption{Scheme of the lock chain. The reference laser (blue) is locked to the Rb spectroscopy. The transfer cavity is locked to the reference laser, which in turn gets a fast feedback from it on the diode current, to narrow the linewidth. The 805 laser (red) is coupled into a fiber EOM, to create sidebands at a tunable distance from the carrier. One of the sidebands is locked to the transfer cavity. This way, by tuning the distance of the sidebands, we can shift the carrier arbitrarily. The main beam remains unmodulated, and is used for trapping as well as to lock the science cavity. The ND filters in the main path and before the PDH lock are placed in a complementary manner, in order to vary the trapping depth, while maintaining the input power to the PDH lock for stable operation. The science laser (green) is locked to the reference using the beat signal of the two.}
 \label{fig:lockChain}
\end{figure*}

Our intra-vacuo high-finesse cavity is aligned horizontally, and its parameters are summarized in \cref{tab:cavity}. Among the series of longitudinal modes of the cavity, we use two for the present experiment.
\begin{enumerate}
 \item A longitudinal mode with wavelength $\approx\SI{805}{\nm}$ is used for a dual purpose
 \begin{enumerate}
  \item locking the science cavity
  \item providing the far-red-detuned optical dipole trap
 \end{enumerate}
 \item Another mode closer to the atomic resonance with wavelength $\approx\SI{780}{\nm}$ (the detuning is $\SI{-90}{\mega\hertz}$ from the $F=2\leftrightarrow3$ transition) is used for monitoring the presence of the atoms within the mode volume through the dispersive shift.
\end{enumerate}
The lock chain for the lasers and the cavity is sketched in \cref{fig:lockChain}. The reference laser is locked directly to rubidium by feedback to the piezo of the ECDL grating. A transfer cavity is locked to this reference.\footnote{An additional subtlety is a much faster feedback loop in the opposite direction: the reference laser is also locked to the transfer-cavity mode. In this case, the latter is used as a narrow-line reference to reduce the linewidth of the reference laser by fast active feedback on the laser diode current.} The 805 laser is a Toptica DL laser without any amplification other than the buildup in the science cavity. It goes through a fiber-based EOM (IXblue NIR-MPX800), and one of the sidebands is locked to the transfer cavity. This modulation allows us to move the science-cavity modes anywhere within the FSR. The science cavity is locked to the 805 laser. The science laser is locked directly to the reference by beating the two together and using either a frequency-to-voltage converter (in the early stage of the present experiment) or a phase-lock-loop (in a later stage) to stabilize the beat frequency. The transmission of the science light through the cavity is monitored by a Thorlabs APD410A/M avalanche photodetector after being separated from the transmitted 805 laser by a Semrock MaxLine Laser-line filter (AHF LL01-780).

The whole experimental sequence is controlled by an ADwin-Pro II realtime process controller, that allows for timing digital and analogue output signals with a precision of $\SI1\us$, and acquire analogue input signals with a resolution down to $\SI{250}\ns$. For defining experimental sequences, we use a Python front-end developed in our group that we are going to present in an upcoming publication.

\section{Determining the effective atom number from the cavity transmittance}
\begin{figure}
 \includegraphics[width=\linewidth]{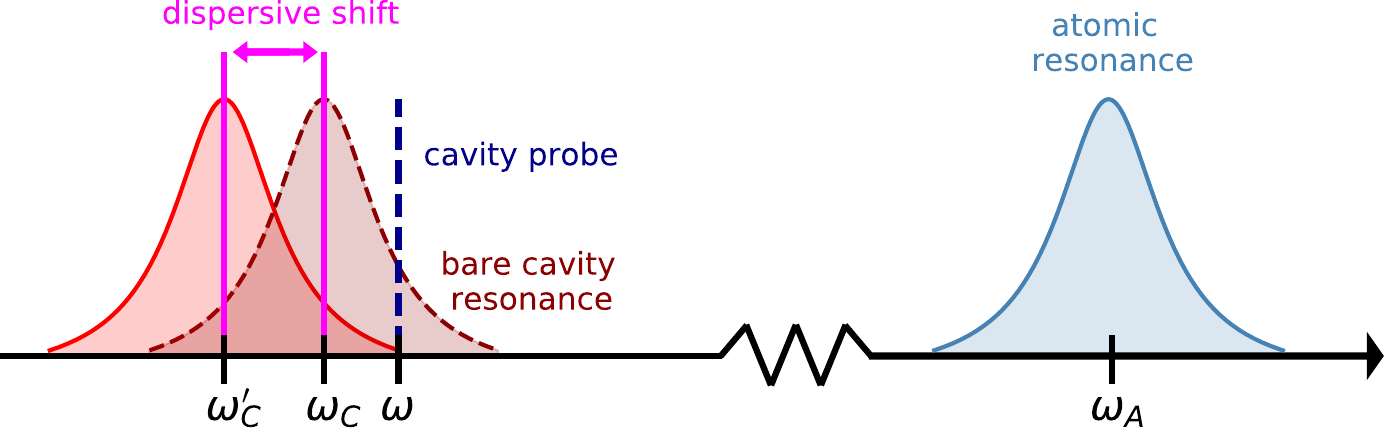}
 \caption{Principle of the detection of atoms. The probe light is red detuned from the atoms by $\SI{-90}{\mega\hertz}$ and blue detuned from the cavity by $\SI{2}{\mega\hertz}$. The empty cavity transmits a fraction of the incoupled light according to the Lorentzian lineshape. Since the atoms introduce a dispersive shift in the cavity, the bare cavity resonance $\omega_C$ is reduced to $\omega_C'$, resulting in a larger detuning from the cavity, and hence the cavity transmittance drops. The ratio of the cavity transmittance with and without atoms can be converted to the number of the atoms which are coupled to the cavity mode (effective atom number) according to \cref{eq:AtomNoInference}.}
 \label{fig:dispersiveShift}
\end{figure}

Denoting the probe frequency, the frequency of the $F=2\leftrightarrow3$ atomic transition, and the bare frequency of the 780-mode of the cavity with $\omega$, $\omega_\text{A}$, and $\omega_\text{C}$, respectively (see \cref{fig:dispersiveShift}), we can define the detunings as crucial parameters in the following, and list their values used in these experiments:
\begin{subequations}
\begin{align}
\omega-\omega_\text{A}&\equiv\Delta_\text{A}=-2\pi\cdot\SI{90}{\mega\hertz},\\
\omega-\omega_\text{C}&\equiv\Delta_\text{C}=2\pi\cdot\SI{2}{\mega\hertz}.
\end{align}
\end{subequations}
In the following, we rely on standard CQED theory. Using that $\Delta_\text{A}\gg\gamma$, where $\gamma=2\pi\cdot\SI{3}{MHz}$ (HWHM) for the Rb D2 line, the dispersive shift caused by the presence of the atoms can be expressed as
\begin{subequations}
\label{eq:AtomNoInference}
\begin{equation}
\omega-\omega_\text{C}'(N_\text{eff})\equiv\Delta_\text{C}'(N_\text{eff})=\Delta_\text{C}-\frac{N_\text{eff}\,g^2}{\Delta_\text{A}},
\end{equation}
where $N_\text{eff}$ is the effective atom number, i.e. the overlap integral of the atomic density distribution and the cavity mode function. Due to the opposite sign of the two detunings, the cavity is further detuned from the resonance by the presence of the atoms. We can define the cavity transmittance as the ratio between the output 780 intensity measured with and without atoms, and express it with the above parameters:
\begin{equation}
\label{eq:Transmittance}
\frac{I(N_\text{eff})}{I_\text{empty cavity}}\equiv\mathcal T(N_\text{eff})=
\frac{\kappa^2+\Delta_\text{C}^2}{\kappa^2+\Delta_\text{C}'(N_\text{eff})^2}.
\end{equation}
\end{subequations}
Hence, by putting together the two equations \labelcref{eq:AtomNoInference}, and using the known CQED parameters, we can conduct a time-resolved measurement of the intra-cavity atom number. This principle is illustrated in \cref{fig:dispersiveShift}.

\section{Intra-cavity dipole trapping}
In the following, we assume that the magnetic transport ends at $t=0$, and the magnetic trap is ramped down over a period of $\SI7\ms$. This is the optimum in our present system, since a quicker ramp causes oscillations in the cavity piezo. The 805 light is present in the cavity at all times, to provide for the cavity lock. Its intensity impinging on the incoupling mirror is adjustable from $\SI{100}{\micro\watt}$ up to $\SI{12}{mW}$.

\begin{figure*}
 \includegraphics[width=0.9\linewidth]{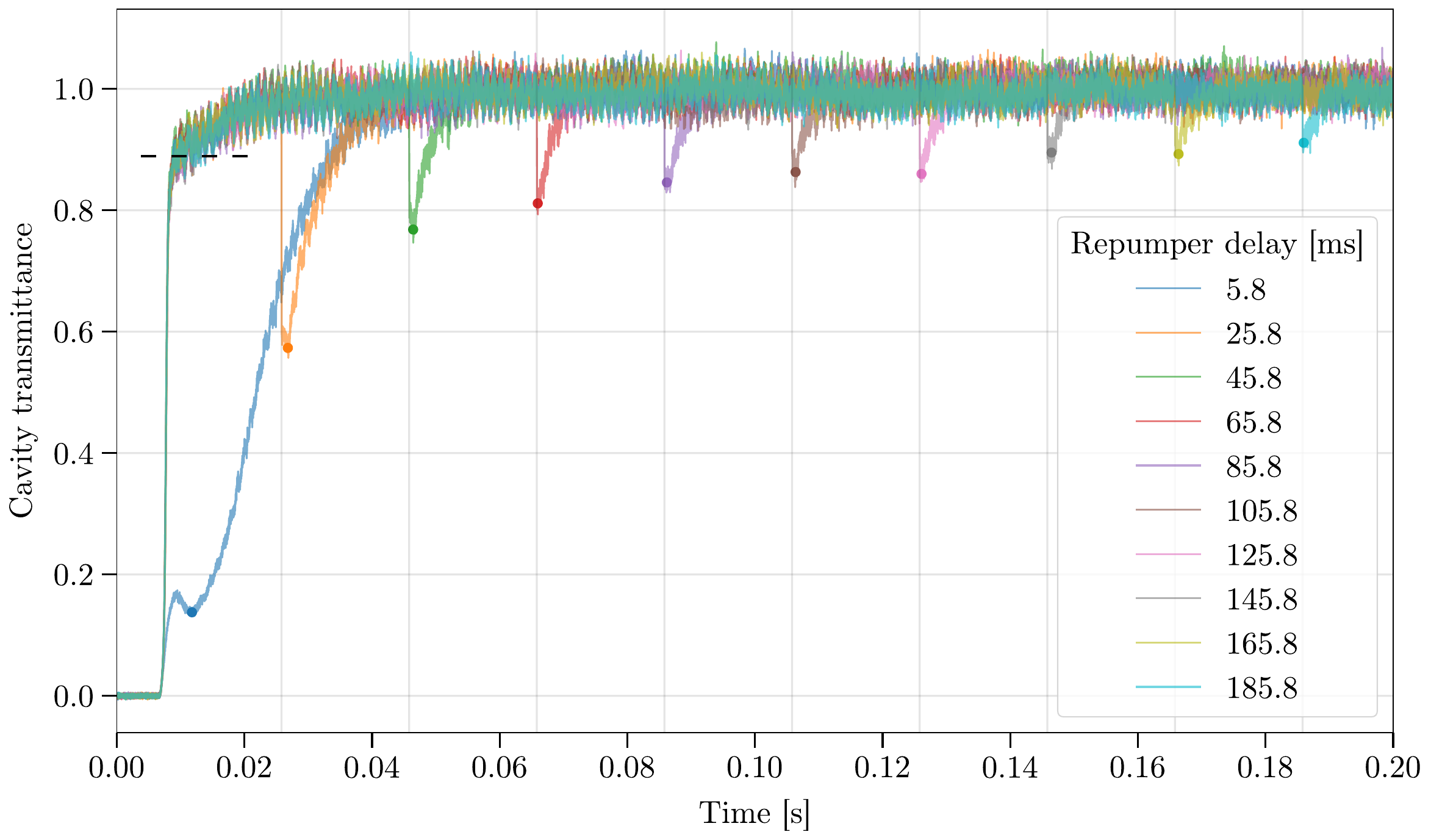}
 \caption{Stroboscopic estimate of the atom number. Each curve is an average of 10 runs and corresponds to a different delay of the repumper exposure. The probe light is open at $\SI{7}{ms}$ in every run. The abrupt rise from $0$ to $\sim0.9$ (black dashed line) in the beginning of all the curves but the blue one corresponds to the timescale of the shutter opening, and to the depumping process to the $F=1$ state. The transmittance does not reach its maximum because of the atoms falling into the interaction volume from the cloud surrounding the mode. When the repumper is turned on, the transmittance drops suddenly, due to the dispersive shift of the atoms which recover from $F=1$ to $F=2$. The averages around the minima at these dips (denoted by coloured dots) are used to determine the effective number of atoms for the exponential fit in \cref{fig:decay}. Concerning the blue curve, the initial fast rise remains below $0.2$, because in this case the repumper is already on, when the probe light enters the cavity, and hence the depumping is disabled. The small dip afterwards can be explained with the trapping effect of the probe light itself (see text). All of the curves, including the blue one, once they reach their minimum, tend to $\mathcal T=1$, as the cavity field and the repumper together heat the atoms out of the trap.}
 \label{fig:stroboscopic}
\end{figure*}

The protocol for characterizing the dipole trap is illustrated in \cref{fig:stroboscopic}. The probe light is switched on right after the switch-off of the magnetic trap at $t=\SI7ms$. The cavity transmittance quickly increases due to the probe light optically pumping atoms from the $F=2$ to the $F=1$ ground state, where their dispersive shift is negligible. From this state, it is possible to bring back the atoms to the strongly-coupled $F=2$ state by the application of a repumper, that we do with increasing time delay.

Upon the switch-on of the repumper, the cavity transmittance drops abruptly, indicating dispersive shift by atoms that remained in the dipole trap in $F=1$ now reappearing in the $F=2$ ground state. Figure \ref{fig:stroboscopic} shows dips from the transmittance $\mathcal T=1$ with different colours corresponding to different times of the repumper switch-on. The transmission then goes back to $\mathcal T=1$ on a time scale of $\approx\SI{20}{\ms}$ because of the evaporation of the atoms that are subject to recoil heating when the repumper is on and there is light in the cavity resonant with the $F=1\leftrightarrow2$ transition. However, this measurement reveals that atoms, in the $F=1$ “dark” ground state, are captured as long as  around $\SI{200}{\ms}$.

Observe that after the very fast initial increase until about $\SI{7}{ms}$, the cavity transmittance reaches 1 only over a period of $\sim\SI{40}\ms$ on those trajectories where the repumper is switched on after this interval (all but the blue curve). This is the effect of the surrounding cloud after the switch-off of the magnetic trap: many atoms that do not get trapped in the cavity mode still transit it during this period. These atoms are in the $F=2$ ground state, so they temporarily contribute to the dispersive shift until they either get depumped into $F=1$ or simply leave the cavity mode. A temporary equilibrium is maintained, which slowly tends towards $\mathcal T=1$, as the buffer depletes. The $\SI{40}{\ms}$ time interval is consistent with our expansion measurements by absorption imaging.

The blue curve, corresponding to the repumper being on from the outset (delay $=\SI{5.8}{\ms}$), exhibits slightly different characteristics in \cref{fig:stroboscopic}. Due to the presence of a strong repumper,  the depumping is inhibited in this case, which leads to a gradual increase of the transmittance as the atoms are heated out of the trap. The small dip at the beginning can be attributed to a rearrangement of the atoms as they all synchronously start to move towards the emerging potential minima at the antinodes of the probe mode red-detuned with respect to the atomic frequency. Approaching the antinodes lead to a higher coupling between the mode and the atoms, which ultimately increases the shift of the cavity resonance, and decreases the transmittance. Even the orange curve (delay $=\SI{25.8}{\ms}$) exhibits this effect, since it blends into the blue one starting from around $t=\SI{30}{\ms}$.

\begin{figure}
 \includegraphics[width=\linewidth]{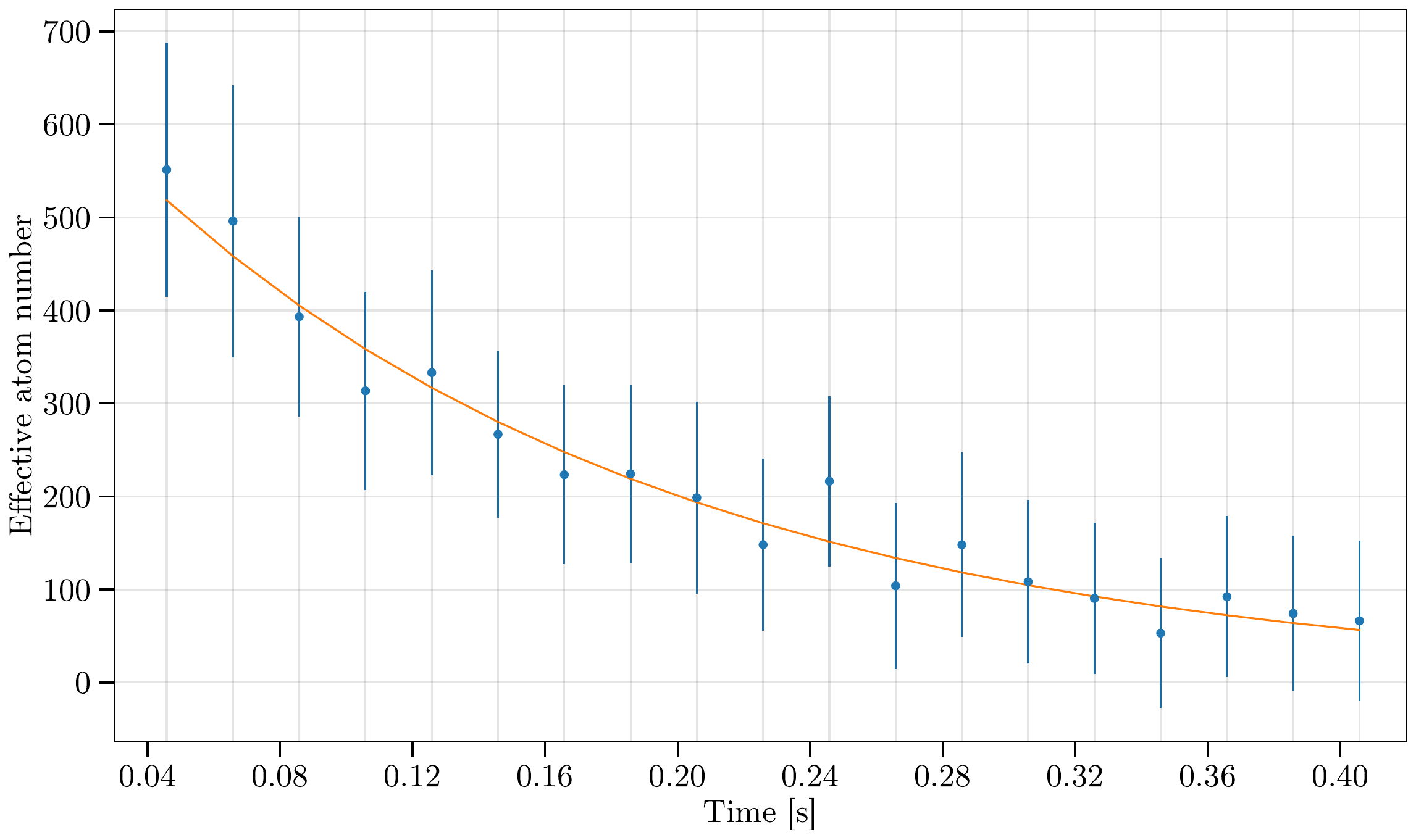}
 \caption{Determining the trapping time. The effective atom number is calculated from the average around the minima of the transmittance curves in \cref{fig:stroboscopic} (coloured dots), utilizing \cref{eq:AtomNoInference}. Only values taken after $\SI{40}{\ms}$ are considered, due to the initial effects of the atomic cloud surrounding the cavity mode. The trapping time acquired from the exponential fit is $\tau=160\pm10\unit{\ms}$}
 \label{fig:decay}
\end{figure}

From the transmittance values determined by the reappearing dispersive shift when the repumper is switched on (identically colored dots on the same figure), we can infer the effective atom number still present in the 805 dipole trap at these time instants, in accordance with \cref{eq:AtomNoInference}. We use only data for which the repumper is switched on after the initial $\SI{40}{\ms}$ period and when the atom number is lower to avoid direct atom-atom interaction effects. The inferred atom numbers are plotted on \cref{fig:decay} as a function of time (i.e. the switch-on time of the repumper). The values fit well on an exponential decay curve, allowing us to extract a single characteristic trapping time $\tau\approx \SI{160}{ms}$. We note that the largest atom number seen for the blue curve at around $t=\SI{9}{ms}$ was $N_\text{eff}=6000 \pm 400$.

\begin{figure}
 \includegraphics[width=\linewidth]{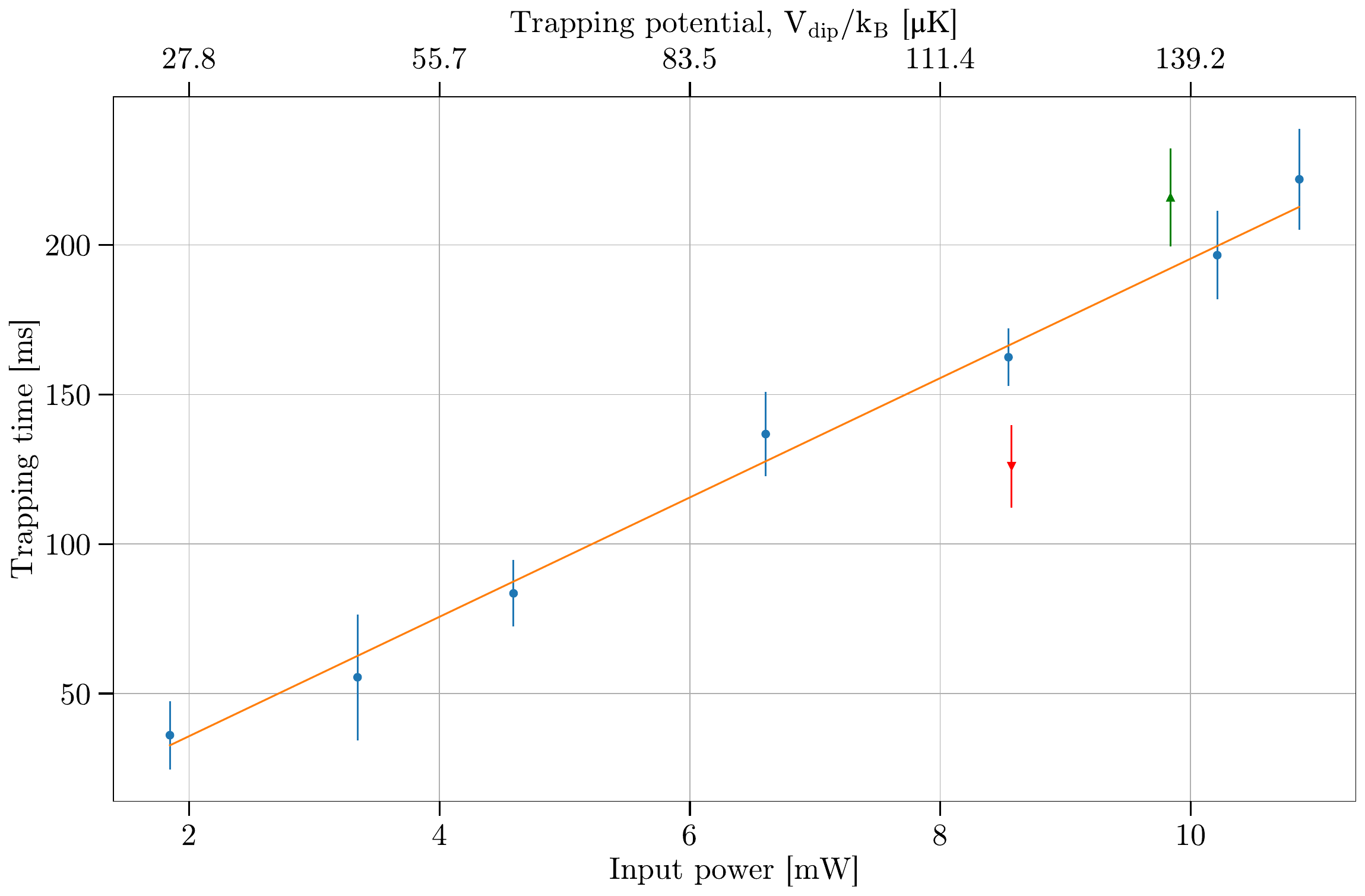}
 \caption{Trapping time as a function of the trapping potential. Each value corresponds to a measurement sequence similar to the one shown in \cref{fig:stroboscopic}, with different input powers of the trapping laser. The measurements for the fitting (linear function, blue points) were taken after optimizing for the vertical transport of the cloud, but before positioning in the horizontal direction. The red point corresponds to the case before any optimization (marked with red triangle on \cref{fig:transportposition}), and the green one was achieved with the optimal vertical and horizontal position of the cloud. There is a significant improvement after each step of optimization.
}
 \label{fig:trappingtime}
\end{figure}

The “smoking-gun” evidence that we have dipole trapping of a \emph{thermal} cloud is presented in \cref{fig:trappingtime}. Here, we vary the intensity of the trapping laser impinging on the cavity. This is achieved by changing the optical density (OD) of the neutral density (ND) filter in the beam path of the 805 light, cf. \cref{fig:lockChain}.\footnote{Another ND filter right before the photodetector of the science cavity PDH lock has to be varied in a complementary way to keep the combined OD constant. A closely constant intensity on this photodetector is needed for the stability of the lock.} As a function of the input power, the trapping time varies in what can be fit with a linear function. This is expected on the basis of the following simple consideration. The atoms with temperature $T$ are sitting in the dipole potential $V_{\mathrm{dip}}$, which is proportional to the trapping laser intensity. The atoms are heated due to diffusion, characterized by the diffusion coefficient $D$. Then the trapping time can be estimated as the ratio of the binding energy and the heating rate, i.e., 
\begin{equation}
    \tau_\mathrm{trap}\approx\frac{V_\mathrm{dip}-k_\mathrm{B}T}{D}
\end{equation}
We assume that the diffusion in our system is dominated by technical noise, which can be expected, since parametric heating is strong when the dipole trap is sustained by a high-finesse cavity, so that $D$ is independent of the intensity.\footnote{In an ideal situation the diffusion originates from the recoil heating induced by the illumination by the trapping laser. Then, assuming also very cold atoms ($k_\mathrm{B}T\ll V_\mathrm{dip}$), since the diffusion is also proportional to the trap laser intensity, the trapping time would exhibit no (or weak) dependence on the intensity.} Then, the intercept of the linear curve gives a good estimate of the temperature of the atomic ensemble, which in our case is $T=2.9\pm\SI{3.8}{\micro\K}$. Even though the error is large, this temperature is significantly lower than what we measure in the magnetic trap ($\gtrsim\SI{10}{\micro K}$). This might indicate that evaporative cooling takes place when we gradually decrease the magnetic trap with a rate decelerating according to a tangent hyperbolic function. To prove this, we need to implement a more precise temperature measurement in the intra-cavity dipole trap.

\section{Optimization of the process}
The most important lever to optimize the trapping is the final position of the magnetic trap, as the magnetically trapped cloud at this point must have very good overlap with the cavity mode. The kinematics of the switch-off of the magnetic trap is also important: here, our experience is the faster the better, however, we cannot go below $\approx\SI{7}{\ms}$, due to the above mentioned instability of the cavity lock caused by very quick changes of strong magnetic fields. Therefore, in the following we concentrate on the final position of the magnetic trap.

\begin{figure}
 \includegraphics[width=\linewidth]{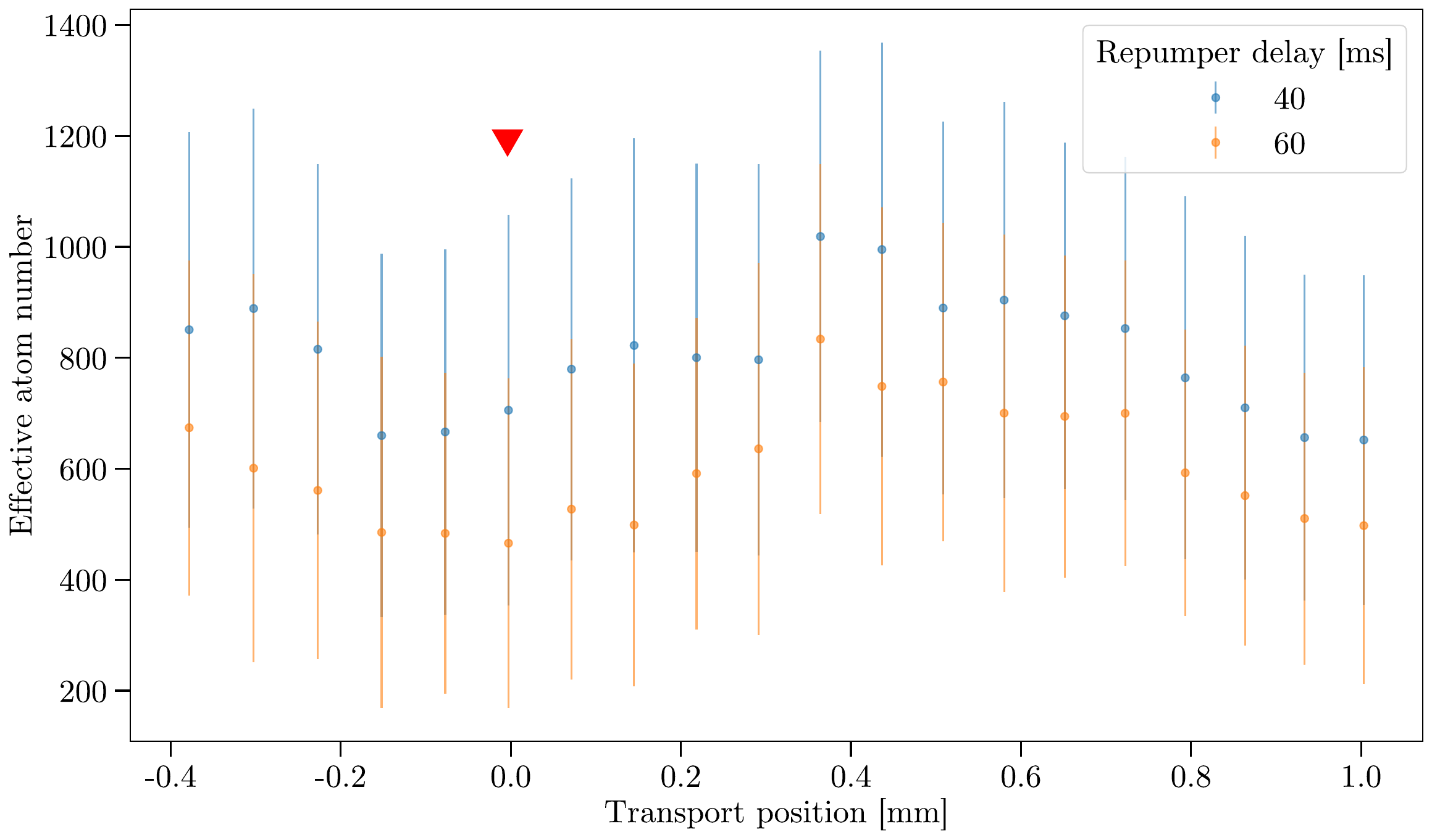}
 \caption{Optimization of the vertical transport position. Before the optimization, the cloud was transported to the geometric center of the cavity ($0$ of the $x$ axis, red triangle). Transporting to a higher position ($\sim\SI{0.36}{\mm}$) results in a slightly higher number of trapped atoms, and also significantly longer trapping time (cf. red point in \cref{fig:trappingtime}).}
 \label{fig:transportposition}
\end{figure}

The vertical position of the magnetic trap can be very easily varied with our dual quadrupole electromagnets, this is what happens also during the magnetic transport. In \cref{fig:transportposition}, we show data of the effective atom number in the dipole trap after $40$ and $\SI{60}{\ms}$ as a function of the final vertical position of the magnetic trap. The atom number exhibits a peak with only moderate significance. However, the trapping time is significantly improved when using the optimum value (by $\sim\SI{36}{\ms}$). This can be seen by comparing the red point in \cref{fig:trappingtime} (which corresponds to the point in \cref{fig:transportposition} marked with red triangle) with the blue point corresponding to the optimum vertical position.

We have performed a similar optimization of the horizontal position, by varying the current \emph{in situ} in the coil pair used for compensating the Earth magnetic field whose axis is perpendicular to the cavity axis in the horizontal plane. This is necessary because due to imperfections of e.g. the winding of the electromagnets, the axis of the quadrupole slightly deviates from the geometrical axis of the setup, making that the center of the magnetically transported cloud misses the cavity mode by $\sim\SI{0.87}{\mm}$. By superimposing a weak homogeneous field on the quadrupole, we can shift the magnetic trap center in this direction, attempting to push it to within the cavity mode. As a function of the current of this coil pair, we cannot observe any significant optimum in the atom number in the dipole trap, however, the trapping time could again be significantly improved (by more than $\sim\SI{20}{\ms}$), as demonstrated by the green point in \cref{fig:trappingtime}.

\begin{figure}
 \includegraphics[width=\linewidth]{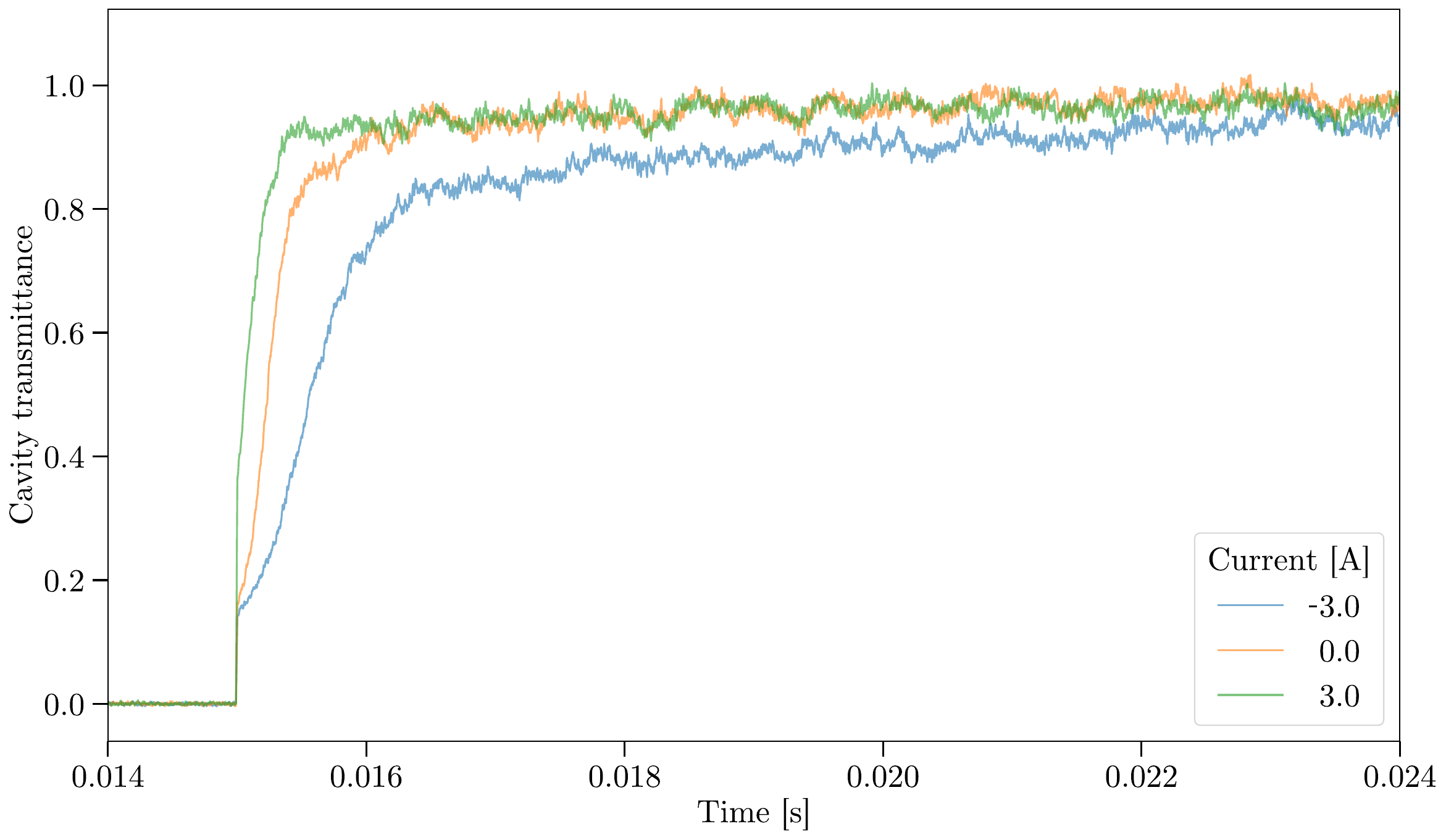}
 \caption{Depumping atoms with different axial magnetic fields. Atoms are transported and released similarly as in the previous measurements, the probe light is turned on at $\SI{15}{\ms}$, and the repumper is not used. Homogeneous magnetic fields are created along the cavity axis, utilizing the coils compensating the stray magnetic fields. The initial rise of the curve is significantly slower when the magnetic field points in a certain direction (blue), than in the case of the opposite direction (green), or when the field is absent (orange). In the absence of magnetic field, or when it is opposite to the polarization of the circularly polarized probe light, $F=2\to1$ depumping can occur, while with a matched magnetic field direction it can be disabled by confining the atoms  into the $(F,m_F)=(2,2)\leftrightarrow(3,3)$ closed cycle. }
 \label{fig:disabledepump}
\end{figure}

As an outlook, we note that it is also possible to control the $F=2\to1$ depumping process by controlling the magnetic field coaxial with the cavity, which we can do with another coil pair of the set used for compensating the magnetic field of the Earth. Since the 780 cavity probe light is circularly polarized, it is in principle possible to completely disable the depumping. For this, during the ramp-down of the magnetic quadrupole trap, we have to ramp up the coaxial homogeneous magnetic field adiabatically. Then, the magnetically trapped atoms that are in the $(F,m_F)=(2,2)$ ground state with respect to the local magnetic field in the quadrupole (which is a prerequisite of purely magnetic trapping) adiabatically rotate their magnetic dipole moment to remain in the same state \emph{with respect to the new, homogeneous field}. If the direction of this coaxial field matches the direction of the polarization of the probe light, then the probe light will cause transitions only in the \emph{closed} cycle $(F,m_F)=(2,2)\leftrightarrow(3,3)$, disabling the depumping. Results shown in \cref{fig:disabledepump} clearly evidence the reinforcement of the transmission suppression by slowing down the depumping process. Depumping is only inhibited, but not perfectly disabled, due to imperfections in the probe polarization and the atomic $m_F$ polarization.

\section{Conclusion}
We have demonstrated the loading of a cold ensemble of $^{87}$Rb atoms from a large magnetostatic quadrupole trap into a small-volume optical dipole trap sustained by a far-red-detuned mode of a high-finesse optical cavity. Up to $10^4$ atoms could be loaded into the cavity mode volume. Here the atoms can be monitored via their dispersive shift on another, closely resonant cavity mode. Hence the optical dipole trapping of a cold ensemble of several hundreds of atoms for up to $\SI{200}\ms$ has been demonstrated. Using a repumper laser, we could demonstrate control over the hyperfine ground state of the atoms.

The presented technique will become a key ingredient for further experiments that can be envisaged in our system, including cavity-assisted electromagnetically induced transparency, microwave-to-optical conversion, and atom-interferometry.

\section*{Acknowledgment}
This research was supported by the Ministry of Culture and Innovation and the National Research, Development and Innovation Office within the Quantum Information National Laboratory of Hungary (Grant No. 2022-2.1.1-NL-2022-00004), and within the ERANET COFUND QuantERA program  (MOCA, 2019-2.1.7-ERA-NET-2022-00041). A. D. acknowledges support from the János Bolyai research scholarship of the Hungarian Academy of Sciences.

\bibliography{CavityDipoleTrap}

\end{document}